%% file: main.tex
\documentclass[runningheads]{llncs}
\usepackage{graphicx}
\usepackage{multirow}
\usepackage{multicol}
\usepackage{booktabs}
\usepackage{amsmath}
\usepackage{amssymb}
\usepackage{hyperref}
\usepackage{marvosym}
\usepackage{threeparttable}

\hypersetup{
    colorlinks=true,
    linkcolor=blue,
    filecolor=magenta,      
    urlcolor=cyan,
    pdftitle={Overleaf Example},
    pdfpagemode=FullScreen,
    }
\usepackage{xcolor}

\hypersetup{hidelinks}

\definecolor{darkergreen}{RGB}{21, 152, 56}

\newcommand{\tabincell}[2]{\begin{tabular}{@{}#1@{}}#2\end{tabular}}  

\begin{document}
%
\title{OpenMedIA: Open-Source Medical Image Analysis Toolbox and Benchmark under Heterogeneous AI Computing Platforms}
%

\author{Jia-Xin Zhuang\inst{1} \and
Xiansong Huang\inst{1} \and
 Yang Yang\inst{1,2} \and
Jiancong Chen\inst{1} \and
Yue Yu\inst{1,4} \and 
Wei Gao\inst{1,3} \and
Ge Li\inst{3} \and
Jie Chen\inst{1,3} \and
Tong Zhang\inst{1}\thanks{Corresponding author}
}
\authorrunning{Zhuang et al.}
\titlerunning{OpenMedIA}
\institute{$^1$Peng Cheng Laboratory, Shenzhen, China \\$^2$
 School of Computer Science and Technology, Harbin Institute of Technology(Shenzhen), China\\ $^3$
 School of Electronic and Computer Engineering, Peking University, China\\ $^4$ 
 National Laboratory for Parallel and Distributed Processing, National University of Defense Technology, China\\
\email{\{zhangt02\}@pcl.ac.cn}}

\maketitle              

\input{sections/abstract}
\input{sections/introduction}
\input{sections/supportedAlgorithm}

\input{sections/benchmark}

\input{sections/conclusion}

\bibliographystyle{splncs04}
\bibliography{egbib}

\end{document}

%% file: sections/abstract.tex
\begin{abstract}
In this paper, we present OpenMedIA, an open-source toolbox library containing a rich set of deep learning methods for medical image analysis under heterogeneous Artificial Intelligence (AI) computing platforms. Various medical image analysis methods, including 2D$/$3D medical image classification, segmentation, localisation, and detection, have been included in the toolbox with PyTorch and$/$or MindSpore implementations under heterogeneous NVIDIA and Huawei Ascend computing systems. To our best knowledge, OpenMedIA is the first open-source algorithm library providing compared PyTorch and MindSpore implementations and results on several benchmark datasets. The source codes and models are available at \url{https://git.openi.org.cn/OpenMedIA}.
\end{abstract}

%% file: sections/introduction.tex
\section{Introduction}

Deep learning has been extensively studied and has achieved beyond human-level performances in various research and application fields~\cite{floridi2020gpt,silver2017mastering,zhang2020shmnet,zhou2021ensembled}. 
To facilitate the development of such Artificial Intelligence (AI) systems, a number of open-sourced deep learning frameworks including TensorFlow~\cite{abadi2016tensorflow}, PyTorch~\cite{paszke2019pytorch}, MXNet~\cite{chen2015mxnet},  MindSpore~\footnote{\url{https://www.mindspore.cn}} have been developed and integrated to various AI hardwares. 
The medical image analysis community also witnessed similar rapid developments and revolutions with deep learning methods developed for medical image reconstruction, classification, segmentation, registration, and detection~\cite{chen2021ellipsenet,cciccek20163d,rueckert2019model,9001020,10.1007/978-3-030-33843-5_2,zhuang2020deep}. 
In particular, MONAI~\footnote{\url{https://monai.io}}, implemented in PyTorch, is a popular open-source toolbox with deep learning algorithms in healthcare imaging. 
Despite its popularity, the accuracy and performance of those algorithms may vary when implemented on different AI frameworks and$/$or various AI hardware. 
Not to mention that several PyTorch or CUDA libraries included in MONAI are not supported by Huawei Ascend and$/$or other NPU computing hardware. 

In this paper, we report our open-source algorithms and AI models library for medical image analysis, OpenMedIA, which have been implemented and verified with various PyTorch and MindSpore AI frameworks with heterogeneous computing hardware.
All the algorithms implemented with the MindSpore framework in OpenMedIA have been evaluated and compared to the original papers and the PyTorch version.  

\input{tables/summary}

Our contributions can be summarized as follows: 

Firstly, we provide an open-source library of recent State-Of-The-Art (SOTA) algorithms in the medical image analysis domain under two deep learning frameworks: PyTorch (with NVIDIA) and MindSpore (with Huawei Ascend). 

Secondly, we conduct bench-marking comparisons of the SOTA algorithms with accuracy and performances. 

Thirdly, we not only open-source the codes but also provide all the training logs and checkpoints under different AI frameworks. In this study, PyTorch is built with NVIDIA GPU, while MindSpore is with Huawei Ascend.

Table~\ref{tab:summary} shows the basic overview of the algorithms included in OpenMedIA. We categorize these algorithms into Classification, Segmentation, Localisation, and Detection tasks for easy understanding and comparison.

%% file: tables/summary.tex
\begin{table*}[ht]
    \caption{A summary of algorithms in our library.}
    \label{tab:summary}
    \centering
    \begin{tabular}{cccc}
        \toprule
       	Task & Algorithm & PyTorch & MindSpore \\
        \midrule
        \multirow{4}{*}{Classification}
        & Covid-ResNet~\cite{farooq2020covid,he2016deep} & \checkmark & \checkmark\\
        & Covid-Transformer~\cite{dosovitskiy2020image} & \checkmark & -\\
        & U-CSRNet~\cite{huang2020bcdata} & \checkmark & \checkmark \\
        & MTCSN~\cite{kong2021multi} & \checkmark & \checkmark\\
        \hline
        \multirow{6}{*}{Segmentation} & 2D-UNet~\cite{hofmanninger2020automatic} & \checkmark & \checkmark \\
        & LOD-Net~\cite{cheng2021learnable} & - & \checkmark \\
        & Han-Net~\cite{he2021hybrid} & \checkmark & \checkmark\\
        & CD-Net~\cite{he2021cdnet} & \checkmark & \checkmark\\
        & 3D-UNet~\cite{tong20173d} & \checkmark & \checkmark \\
        & UNETR~\cite{hatamizadeh2022unetr} & \checkmark & -\\
        \hline 
        \multirow{2}{*}{Localization} & WeaklyLesionLocalisation~\cite{yang2021towards} & \checkmark & - \\
        & TS-CAM~\cite{gao2021ts} & \checkmark & -\\
        \hline
        \multirow{5}{*}{Detection} &  LungNodules-Detection~\cite{setio2017validation} & - & \checkmark \\
        & Covid-Detection-CNN~\footnote{\url{https://github.com/ultralytics/yolov5}} & \checkmark & -\\
        & Covid-Detection-Transformer~\cite{carion2020end} & \checkmark & -\\
        & EllipseNet-Fit~\cite{chen2021ellipsenet,sinclair2018human} & - & \checkmark \\
        & EllipseNet~\cite{chen2021ellipsenet} & \checkmark & - \\
        \bottomrule
    \end{tabular}
\end{table*}

%% file: sections/supportedAlgorithm.tex
\section{Algorithms}
This study summarises seventeen SOTA algorithms from medical image classification, segmentation, localisation, and detection tasks. Eight of them are implemented by both MindSpore and PyTorch. We will continue updating the OpenMedIA library in the next few years. In this section, we will briefly introduce the selected algorithms.

\subsection{Medical image classification}
Four well-known methods are introduced for this task.  It should be noted that the current open-source codes and models are with 2D classification settings. \\

\noindent{\textbf{Covid-ResNet.}} The contributions of Covid-ResNet~\footnote{PyTorch: \url{https://git.openi.org.cn/OpenMedIA/Covid-ResNet.Pytorch}}~\footnote{MindSpore: \url{https://git.openi.org.cn/OpenMedIA/Covid-ResNet.Mindspore}} are listed as follows:
	\begin{itemize}
		\item ResNet~\cite{he2016deep} was proposed in 2015 and became one of the most famous Convolutional Neural Networks(CNN) in deep learning. 
		\item It uses a residual learning framework to ease the training of a deeper network than previous work and shows promising performance. 
		\item An early CNN model built for COVID-19 CT image classification.
	\end{itemize}

\noindent{\textbf{Covid-Transformer.}} The contributions of Covid-Transformer~\footnote{PyTorch: \url{https://git.openi.org.cn/OpenMedIA/Covid-Transformer.Pytorch}} are listed as follows:
	\begin{itemize}
		\item ViT~\cite{dosovitskiy2020image} was inspired by the success of the transformer in Natural Language Process(NLP) and proposed for Computer Vision in 2020. 
		\item Unlike a convolutional network like ResNet, which includes a convolution structure to extract features, ViT consists of self-attention. It doesn't introduce any image-specific inductive biases into the architecture. ViT interprets an image as a sequence of patches and processes it by a pure encoder, which shows comparable performance to CNNs.
	\end{itemize}

\noindent{\textbf{U-CSRNet.}} The contributions of U-CSRNet~\footnote{PyTorch: \url{https://git.openi.org.cn/OpenMedIA/U-CSRNet.Pytorch}}~\footnote{MindSpore: \url{https://git.openi.org.cn/OpenMedIA/U-CSRNet.Mindspore}} are listed as follows:
	\begin{itemize}
		\item U-CSRNet~\cite{huang2020bcdata} add transpose convolution layers after its backend so that the final output probability map can be identical to the input's resolution. And modify the output of CSRNet and U-CSRNet from one channel to two channels to represent the two kinds of tumor cells.
	\end{itemize}

\noindent{\textbf{MTCSN.}} The contributions of MTCSN~\footnote{PyTorch: \url{https://git.openi.org.cn/OpenMedIA/MTCSN.Pytorch}}~\footnote{MindSpore: \url{https://git.openi.org.cn/OpenMedIA/MTCSN.Mindspore}} are listed as follows:
	\begin{itemize}
		\item MTCSN is mainly used to evaluate the definition of the capsule endoscope. 
		\item It is an end-to-end evaluation method. 
		\item MTCSN uses the structure of ResNet in the encoding part and designs two multi-task branches in the decoding part. Namely, the classification branch for the availability of image definition measurement and the segmentation branch for tissue segmentation generates interpretable visualization to help doctors understand the whole image.
	\end{itemize}

\subsection{Medical image segmentation}
Most methods are designed for 2D segmentation tasks. We also include 3D-UNet and UNETR for 3D segmentation tasks for further research.\\

\noindent{\textbf{2D-UNet.}} The contributions of 2D-UNet~\footnote{PyTorch: \url{https://git.openi.org.cn/OpenMedIA/2D-UNet.Pytorch}}~\footnote{MindSpore: \url{https://git.openi.org.cn/OpenMedIA/2D-UNet.Mindspore}} are listed as follows:
	\begin{itemize}
		\item 2D-UNet~\cite{hofmanninger2020automatic} was proposed in 2015 as a type of neural network directly consuming 2D images. 
		\item The U-Net architecture achieves excellent performance on different biomedical segmentation applications. Without solid data augmentations, it only needs very few annotated images and has a reasonable training time.
	\end{itemize}
	
\noindent{\textbf{LOD-Net.}} The contributions of LOD-Net~\footnote{MindSpore: \url{https://git.openi.org.cn/OpenMedIA/LOD-Net.Mindspore}} are listed as follows:
	\begin{itemize}
		\item LOD-Net~\cite{cheng2021learnable} is mainly used in the task of polyp segmentation. 
		\item It is an end-to-end segmentation method. Based on the mask R-CNN architecture, the parallel branch learns the directional derivative of the pixel level of the feature image, measures the gradient performance of the pixels on the image with the designed strategy, and is used to sort and screen out the possible boundary regions. 
		\item The directional derivative feature is used to enhance the features of the boundary regions and, finally, optimize the segmentation results
	\end{itemize}

\noindent{\textbf{HanNet.}} The contributions of HanNet~\footnote{PyTorch: \url{https://git.openi.org.cn/OpenMedIA/HanNet.Pytorch}}~\footnote{MindSpore: \url{https://git.openi.org.cn/OpenMedIA/HanNet.Mindspore}} are listed as follows:
	\begin{itemize}
		\item HanNet~\cite{he2021hybrid} proposes a hybrid-attention nested UNet for nuclear instance segmentation, which consists of two modules: a hybrid nested U-shaped network (H-part) and a hybrid attention block (A-part). 
	\end{itemize}

\noindent{\textbf{CDNet.}} The contributions of CDNet ~\footnote{PyTorch: \url{https://git.openi.org.cn/OpenMedIA/CDNet.Pytorch}}~\footnote{MindSpore: \url{https://git.openi.org.cn/OpenMedIA/CDNet.Mindspore}} are listed as follows:
	\begin{itemize}
		\item CDNet~\cite{he2021cdnet} propose a novel centripetal direction network for nuclear instance segmentation.
		\item The centripetal feature is defined as a class of adjacent directions pointing to the core center to represent the spatial relationship between pixels in the core. Then, these directional features are used to construct a directional difference graph to express the similarity within instances and the differences between instances. 
		\item This method also includes a refining module for direction guidance. As a plug-and-play module, it can effectively integrate additional tasks and aggregate the characteristics of different branches.
	\end{itemize}

\noindent{\textbf{3D-UNet.}} The contributions of 3D-UNet~\footnote{PyTorch: \url{https://git.openi.org.cn/OpenMedIA/3D-UNet.mindspore}}~\footnote{MindSpore: \url{https://git.openi.org.cn/OpenMedIA/3D-UNet.Pytorch}} are listed as follows:
	\begin{itemize}
		\item 3D-UNet~\cite{cciccek20163d} was proposed in 2016, it is a type of neural network that directly consumes volumetric images. 
		\item 3D-UNet extends the previous u-net architecture by replacing all 2D operations with their 3D counterparts. 
		\item The implementation performs on-the-fly elastic deformations for efficient data augmentation during training. It is trained end-to-end from scratch, i.e., no pre-trained network is required.
	\end{itemize}

\noindent{\textbf{UNETR.}} The contributions of UNETR~\footnote{PyTorch: \url{https://git.openi.org.cn/OpenMedIA/Transformer3DSeg}} are listed as follows:
	\begin{itemize}
		\item UNETR~\cite{hatamizadeh2022unetr} was inspired by the success of transformers in NLP and proposed to use a transformer as the encoder to learn sequence representation of the input 3D Volume and capture multi-scale information. 
		\item With the help of a U-shape network design, the model learns the final semantic segmentation output.
	\end{itemize}

\subsection{Weakly supervised image localisation}
Two weakly supervised medical image localisation methods are included in OpenMedIA. Both methods are designed for image localisation tasks with generative adversarial network (GAN) and class activation mapping (CAM) settings. It worth noting that WeaklyLesionLocalisation~\cite{yang2021towards} also supports weakly supervised lesion segmentation. In this work, the comparisons of lesion segmentation were not included. \\

\noindent{\textbf{WeaklyLesionLocalisation.}} The contribution of WeaklyLesionLocalisation~\footnote{PyTorch: \url{https://git.openi.org.cn/OpenMedIA/WeaklyLesionLocalisation}} are listed as follows:
	\begin{itemize}
		\item It~\cite{yang2021towards} proposed a data-driven framework supervised by only image-level labels. 
		\item The framework can explicitly separate potential lesions from original images with the help of a generative adversarial network and a lesion-specific decoder.
	\end{itemize}

\noindent{\textbf{TS-CAM.}} The contribution of TS-CAM~\footnote{PyTorch: \url{https://git.openi.org.cn/OpenMedIA/TS-CAM.Pytorch}} are listed as follows:

	\begin{itemize}
		\item It~\cite{gao2021ts} introduces the token semantic coupled attention map to take full advantage of the self-attention mechanism in the visual transformer for long-range dependency extraction. 
		\item TS-CAM first splits an image into a sequence of patch tokens for spatial embedding, which produce attention maps of long-range visual dependency to avoid partial activation. 
		\item TS-CAM then re-allocates category-related semantics for patch tokens, making them aware of object categories. TS-CAM finally couples the patch tokens with the semantic-agnostic attention map to achieve semantic-aware localisation
	\end{itemize}

\subsection{Medical image detection}
We include seven popular SOTA detection methods in this library and trained them with various medical image detection tasks. \\

\noindent{\textbf{LungNodules-Dectection.}} The contribution of LungNodules-Dectection~\footnote{MindSpore: \url{https://git.openi.org.cn/OpenMedIA/LungNodules-Detection.MS}} are listed as follows:
	\begin{itemize}
		\item It use CenterNet~\cite{zhou2019objects} as the backbone to detect Lung disease. 
		\item CenterNet is a novel practical anchor-free method for object detection, which detects and identifies objects as axis-aligned boxes in an image. 
		\item The detector uses keypoint estimation to find center points and regresses to all other object properties, such as size, location, orientation, and even pose. 
		\item In nature, it’s a one-stage method to simultaneously predict the center location and boxes with real-time speed and higher accuracy than corresponding bounding box-based detectors.
	\end{itemize}

\noindent{\textbf{LungNodule-Detection-CNN.}} The contribution of LungNodule-Detection-CNN~\footnote{PyTorch: \url{https://git.openi.org.cn/OpenMedIA/LungNodule-Detection-CNN.Pytorch}} are listed as follows:
	\begin{itemize}
		\item It use Yolov5~\footnote{\url{https://github.com/ultralytics/yolov5}} the single stage method to detect lung nodules in the Lung. 
		\item Yolov5 is an object detection algorithm that divides images into a grid system. Each cell in the grid is responsible for detecting objects within itself.
	\end{itemize}

\noindent{\textbf{LungNodule-Detection-Transformer.}} The contribution of LungNodule-Detection-Transformer~\footnote{PyTorch: \url{https://git.openi.org.cn/OpenMedIA/LungNodule-Detection-Transformer.Pytorch}} are listed as follows:
	\begin{itemize}
		\item It use DeTR~\cite{carion2020end} to detect the location of the Lung Nodules. 
		\item DeTR is the end-to-end object detection framework based on the transformer, showing promising performance.
	\end{itemize}

\noindent{\textbf{EllipseNet-fit.}} The contribution of EllipseNet-fit~\footnote{MindSpore: \url{https://git.openi.org.cn/OpenMedIA/EllipseFit.Mindspore}} are listed as follows:
	\begin{itemize}
		\item EllipseNet-fit~\cite{chen2021ellipsenet,sinclair2018human} developed a segmentation and ellipse fit network for automatically measuring the fetal head circumference and biparietal diameter. 
		\item Compared to the fetal head ellipse detection, fetal echocardiographic measurement is challenged by the moving heart and shadowing artifacts around the fetal sternum.
	\end{itemize}

\noindent{\textbf{EllipseNet.}} The contribution of EllipseNet~\footnote{PyTorch: \url{https://git.openi.org.cn/OpenMedIA/EllipseNet}} are listed as follows:
	\begin{itemize}
		\item EllipseNet~\cite{chen2021ellipsenet} presents an anchor-free ellipse detection network, namely EllipseNet, which detects the cardiac and thoracic regions in ellipses and automatically calculates the cardiothoracic ratio and cardiac axis for fetal cardiac biometrics in 4-chamber view. 
		\item The detection network detects each object's center as points and simultaneously regresses the ellipses’ parameters. A rotated intersection-over-union loss is defined to further regulate the regression module.
	\end{itemize}

%% file: sections/benchmark.tex
\section{Benchmarks and results}
Algorithms with PyTorch version were implemented and tested on NVIDIA GeForce 1080/2080Ti and Tesla V100~\footnote{\url{https://www.nvidia.com/en-us/data-center/v100/}}. Algorithms with the MindSpore version were trained and evaluated on Huawei Ascend 910~\footnote{\url{https://e.huawei.com/en/products/cloud-computing-dc/atlas/ascend-910}}. Huawei Ascend 910 was released in 2019, which reports twice the performance of rival NVIDIA's Telsa V100~\footnote{\url{https://www.jiqizhixin.com/articles/2019-08-23-7}}. 

\subsection{Datasets}
We conduct seventeen experiments on ten medical image datasets to evaluate different implementations of deep medical image analysis algorithms. 
For 2D classification tasks, the Covid-19 CT image dataset~\footnote{\url{https://covid-segmentation.grand-challenge.org}}, noted as Covid-Classification in this paper, BCData~\cite{huang2020bcdata} and Endoscope~\cite{kong2021multi} datasets were included. For 2D image segmentation tasks, besides the COVID lesion segmentation in Covid-19 CT image dataset $^{32}$ noted as Covid-Segmentation, we also generated a lung segmentation based on this dataset and noted as 2D-Lung-Segmentation~~\footnote{\url{https://git.openi.org.cn/OpenMedIA/2D-UNet.Pytorch/datasets}}. Besides, ETIS-LaribPolypDB~\cite{siegel2020colorectal} and MoNuSeg~\cite{kumar2017dataset} were also used in this study. For 3D image segmentation tasks, we adopted MM-WHS~\cite{zhuang2013challenges} dataset. For 2D image detection tasks, both LUNA16~\footnote{\url{https://luna16.grand-challenge.org/}} and fetal four chamber view ultrasound (FFCV) dataset~\cite{chen2021ellipsenet} were evaluated. Some of the datasets with public copyright licenses have been uploaded to OpenMedIA.

\input{tables/sota_reimplements}

\subsection{Metrics}
For 2D classification task, accuracy(Acc)~\cite{scikit-learn} and F1~\cite{scikit-learn} are used to evaluate the classification accuracy. For 2D/3D segmentation tasks, Dice score~\cite{cciccek20163d} is used to measure the segmentation accuracy. For weakly supervised lesion localisation and segmentation tasks, AUC~\cite{ling2003auc} and Dice score~\cite{cciccek20163d} are used for quantitative comparisons. For 2D detection tasks, mAP~\cite{pascal-voc-2007} , AUC~\cite{ling2003auc} and Dice score are used for evaluation. 

\subsection{Evaluation on image classification}
In Table~\ref{tab:sota}, Rows 1-4 show the accuracy of four image classification algorithms with different implementations. 
It can be seen that CNN achieves better performance than Transformer backbones, which indicates that Transformers are more difficult to train and usually require large training samples.
Experimental results show that the algorithms implemented with PyTorch and MindSpore 
achieve the same accuracy. It should be noted that the reported classification algorithms and results are implemented and verified with 2D scenarios.  
\subsection{Evaluation on image segmentation}
Both 2D and 3D segmentation methods are evaluated, and their results are shown in Rows 5-10 of Table~\ref{tab:sota}. For 2D segmentation tasks shown in Rows 5-8, our implementation achieves comparable or even better performance than the reported results. For 3D segmentation tasks shown in Rows 9-10, 3D-Net is based on CNN architecture, while UNETR uses Transformer as the backbone. It is also worth noting that the 2D algorithms achieve similar results in both MindSpore and PyTorch frameworks, 
while the 3D-UNet algorithm implemented in MindSpore shows a 0.03 accuracy drop compared to the PyTorch re-implementation, which indicates that the MindSpore version could be further optimized in the future, especially on the 3D data augmentation libraries.

\subsection{Evaluation on weakly supervised image localisation}
For medical image localisation, two well-known weakly supervised methods methods are reported in Rows 11-12. Following the settings reported by~\cite{yang2021towards}, both methods are evaluated on the Covid-Segmentation datasets. The implementations of the MindSpore framework will be added in the future.

\subsection{Evaluation on image detection}
For image detection, the results on the Rows 13-14 show that networks based on both CNN and Transformer backbones apply to medical image detection tasks. YOLOv5m are used for CNN based network. In most medical image detection scenarios, rotated ellipses and$/$or bounding boxes are more suitable considering the rotating and ellipse-shaped targets. In this study, we include EllipseNet~\cite{chen2021ellipsenet} and its variant~\cite{sinclair2018human} in OpenMedIA. Rows 16-17 show that EllipseNet delivers the highest accuracy for rotated ellipse detection tasks.

\input{figures/speed.tex}
\subsection{Evaluation on time efficiency}
To evaluate the time efficiency of different implementations with PyTorch and MindSpore, we select one method from each category and evaluate their inference time as shown in Figure~\ref{fig:speed}. For fair comparisons, all the algorithms implemented with PyTorch were assessed on a single NVIDIA Tesla V100, while those implemented with MindSpore were tested on a single Huawei Ascend 910. Figure~\ref{fig:speed} shows the compared time efficiency results. It can be seen that all the tested algorithms with MindSpore environment are more time-efficient than that with PyTorch settings. The time cost of the former is less than 1/3 of that of the latter, which confirms the time efficiency advantage of MindSpore and Huawei Ascend 910 as reported in.

%% file: tables/sota_reimplements.tex
\begin{table*}[ht]
    \caption{Comparative algorithm accuracy. Experimental results reported in the Original columne are directly taken from the paper. We use - to represent lack of implementations or reports. For instance, the original results of Covid-Transformer and TS-CAM are not included because we first conduct such studies. The Re-implement column reports the results of our implementations with PyTorch and$/$or MindSpore AI frameworks.}
    \label{tab:sota}
    \resizebox{\linewidth}{!}{ 
    \begin{tabular}{ccccc}
        \toprule
         \multirow{2}{*}{Algorithm} & \multirow{2}{*}{Dataset} & \multirow{2}{*}{Metric} & \multicolumn{2}{c}{Performance}\\
         & & & Original & \tabincell{c}{Re-implement\\(PyTorch/MindSpore)} \\
         \midrule
         Covid-ResNet~\cite{farooq2020covid} & Covid-Classification & Acc & 0.96 & 0.97/0.97\\ 
         Covid-Transformer~\cite{dosovitskiy2020image} & Covid-Classification & Acc & - & 0.87/- \\     
         U-CSRNet~\cite{huang2020bcdata} & BCData & F1 & 0.85 & 0.85/0.85\\
         MTCSN~\cite{kong2021multi} & Endoscope & Acc & 0.80 & 0.80/0.75\\ 
		\hline
		 2D-UNet~\cite{hofmanninger2020automatic} & 2D-Lung-Segmentation & Dice & - & 0.76/0.76\\ 
		 LOD-Net~\cite{cheng2021learnable} & ETIS-LaribPolypDB & Dice & 0.93 & -/0.91\\ 
		 Han-Net~\cite{he2021hybrid} & MoNuSeg & Dice & 0.80 & 0.80/0.79\\ 
         CD-Net~\cite{he2021cdnet} & MoNuSeg & Dice & 0.80 & 0.80/0.80\\ 
         3D-UNet~\cite{tong20173d} & MM-WHS & Dice & 0.85 & 0.88/0.85\\ 
         UNETR~\cite{hatamizadeh2022unetr} & MM-WHS & Dice & - & 0.84/-\\
        \hline
         WeaklyLesionLocalisation~\cite{yang2021towards} & Covid-Segmentation & \tabincell{c}{AUC} & \tabincell{c}{0.63} & \tabincell{c}{0.63/-} \\
         TS-CAM~\cite{gao2021ts} & Covid-Segmentation & AUC & - & 0.50/-\\
          \hline
         LungNodules-Detection~\cite{setio2017validation} & Luna16 & mAP & - & -/0.51\\ 
         LungNodule-Detection-CNN & Luna16 & mAP & - & 0.72/-\\
         LungNodule-Detection-Transformer~\cite{carion2020end} & Luna16 & mAP & - & 0.72/-\\
         EllipseNet-Fit~\cite{chen2021ellipsenet,sinclair2018human} & FFCV & Dice & 0.91 & 0.91/0.91\\ 
         EllipseNet~\cite{chen2021ellipsenet} & FFCV & Dice & 0.93 & 0.93/-\\ 
         \bottomrule
    \end{tabular}
    }
\end{table*}

%% file: figures/speed.tex
\begin{figure}[ht]
    \centering
    \includegraphics[width=0.7\linewidth]{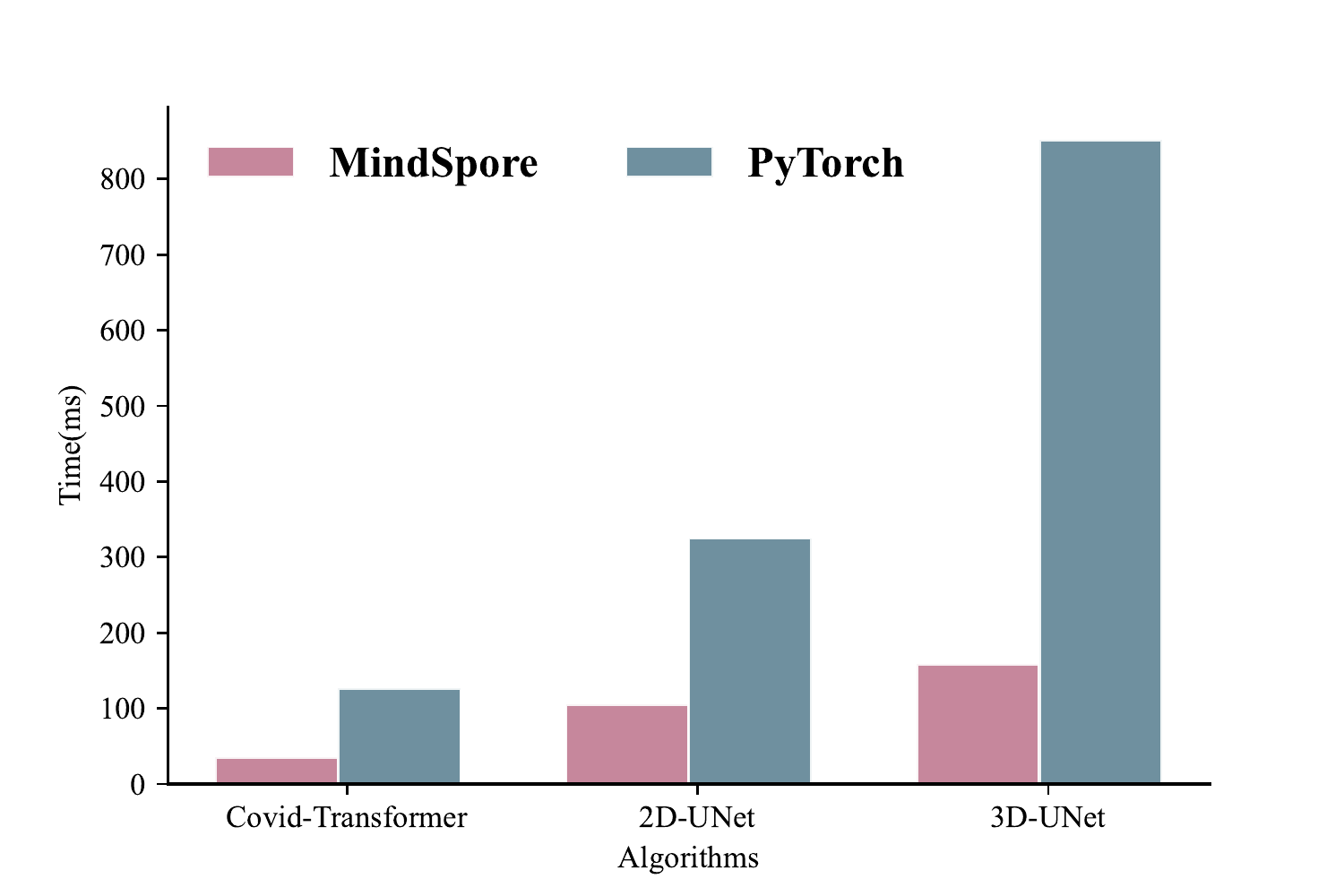}
    \caption{Compared running time of three methods training with different depp learning framework. From left to right in sequence, the methods are Covid-Transformer, 2D-UNet and 3D-UNet.}
    \label{fig:speed}
\end{figure}

%% file: sections/conclusion.tex
\section{Conclusions and future works}
This paper introduces OpenMedIA, an open-source heterogeneous AI computing algorithm library for medical image analysis. We summarise the contributions of a set of SOTA deep learning methods and re-implement them with both PyTorch and MindSpore AI frameworks. This work not only reports the model training/inference accuracy, but also includes the performances of the algorithms with various NVIDIA GPU and Huawei Ascend NPU hardware. The OpenMedIA aims to boost AI development in medical image analysis domain with open-sourced easy implementations, bench-marking models and the open-source computational power. More SOTA algorithms under various AI platforms, such as self-supervised methods will be added in future. 

\section*{Acknowledgement}
The computing resources of Pengcheng Cloudbrain are used in this research. We acknowledge the support provided by OpenI Community (\url{https://git.openi.org.cn}). 